\documentclass[12pt,a4paper]{article}

\usepackage[a4paper,margin=1in]{geometry}
\usepackage{setspace}
\newcommand{\anon}{1}
\usepackage{amsmath,amssymb,amsfonts,mathtools}
\mathtoolsset{showonlyrefs=true}
\usepackage{bm}
\usepackage{mathrsfs}
\usepackage{graphicx}
\usepackage[export]{adjustbox}
\usepackage{booktabs}
\usepackage{array}
\usepackage{multirow}
\usepackage{threeparttable}
\usepackage{tabularx}
\usepackage{siunitx}
\usepackage{longtable}
\usepackage{pdflscape}
\usepackage{rotating}
\usepackage{float}
\usepackage{placeins}
\usepackage{caption}
\usepackage{subcaption}
\usepackage{algorithm}
\usepackage{appendix}

\usepackage[T1]{fontenc}
\usepackage[utf8]{inputenc}
\usepackage{lmodern}

\usepackage[round]{natbib}
\usepackage{xurl}
\usepackage{hyperref}

\usepackage{doi}
\usepackage{url}
\usepackage{nicefrac}

\usepackage{xcolor}

\makeatletter
\def\fps@figure{htbp}
\def\fps@table{htbp}
\makeatother

\begin{document}
\def\spacingset#1{\renewcommand{\baselinestretch}%
{#1}\small\normalsize} \spacingset{1}

\if1\anon
{
  \title{\bf Dynamic Inference in Term Structure Models with Unspanned Latent Risks}
  \author{Tomasz Dubiel-Teleszynski \\
	Liechtenstein Business School, University of Liechtenstein\\
	Fürst-Franz-Josef-Strasse, 9490, Vaduz, Liechtenstein \\
	\texttt{tomasz.dubiel@uni.li} \\
	and \\
	Konstantinos Kalogeropoulos \\
	Department of Statistics, London School of Economics\\
	Houghton Street, London, WC2A 2AE, United Kingdom \\
	\texttt{k.kalogeropoulos@lse.ac.uk} \\
	and \\
	Nikolaos Karouzakis \\
	Alba Graduate Business School, The American College of Greece\\
	6-8 Xenias Str, 115 28 Athens, Greece \\
	\& University of Sussex Business School, Brighton, United Kingdom \\
	\texttt{nkarouzakis@alba.acg.edu}}
  \maketitle
} \fi

\if0\anon
{
  \bigskip
  \begin{center}
    {\LARGE\bf Dynamic Inference in Term Structure Models with Unspanned Latent Risks}
\end{center}
  \medskip
} \fi

\begin{abstract}
We propose a parsimonious class of arbitrage-free, yields-only dynamic term structure models (DTSMs) with unspanned latent risks. To enable sequential estimation and forecasting, we develop a Sequential Monte Carlo framework that combines particle learning for static parameters with Kalman filter updates for latent states, yielding joint posterior inference and predictive distributions that account for both parameter and state uncertainty. We use this framework to assess the out-of-sample statistical and economic value of bond return predictability from the perspective of a Bayesian investor. Empirically, we find that unspanned latent factors contain predictive information beyond that embedded in the yield curve, improving out-of-sample forecasting performance relative to standard benchmark models. These gains translate into economically meaningful utility improvements across a range of portfolio settings. Finally, we show that the hidden component of the slope-related risk factor is countercyclical and associated with real economic activity, suggesting that the latent factors capture economically relevant variation not directly reflected in yields.
\end{abstract}

\noindent%
\vfill

\newpage
\spacingset{1.8} 

\section{Introduction}\label{Intro}

The presence of information not spanned by the yield curve, yet relevant for predicting bond excess returns, has attracted considerable attention in the term structure literature. A large body of empirical work (e.g., \cite{Cochrane05, Ludvigson09, Cooper09, Duffee11, Joslin14, Cieslak15, CoroneoGiannoneModugno2016, Gargano19, BauerRudebusch2020, Bianchi20, LiSarnoZinna2024}) documents both statistical predictability and, in some cases, economically significant gains from macro-based predictors. This suggests that the current cross section of yields may not fully span all relevant predictive information for bond risk premia. Recent work also emphasizes the role of persistent low-frequency components in inflation and equilibrium real rates for explaining Treasury yields and term premia \citep{BauerRudebusch2020}.

Other evidence, however, suggests that these findings may be sensitive across samples and forecasting exercises. \cite{Bauer16} and \cite{BauerHamilton18} show that the predictive content of macroeconomic variables is often weak and unstable, particularly out-of-sample, where augmented models do not always outperform simpler yield-based benchmarks. Similar conclusions arise in the context of unspanned macro-finance DTSMs \citep{Giacoletti21}. In addition, the economic value of such predictability appears sensitive to data availability: studies using real-time macroeconomic information (e.g., \cite{Ghysels18, WanFulopLi2022}) find substantially weaker gains relative to analyses based on revised data, raising concerns about the robustness of macro-based predictability in practice. 

An alternative strand of the literature relaxes the view that the current cross section of yields spans all relevant predictive information. \citet{Joslin14} develop a macro-finance DTSM with unspanned macro risks, while \citet{Duffee11} develops Gaussian term structure models in which economically important return-predictive variation may be hidden from contemporaneous yields. More recently, \citet{BauerRudebusch2020} document that time-varying long-run macroeconomic trends can materially improve yield forecasts and term-premium measurement.

We propose a class of arbitrage-free, yields-only DTSMs with unspanned latent risks, designed to capture predictive information not fully reflected in the current yield curve. Our framework combines the canonical observed-factor setup of \citet{Joslin11} with the unspanned-risk perspective of \citet{Joslin14}, but replaces observable macroeconomic factors with an additional latent component inferred directly from yields under identification-preserving restrictions. Specifically, the model decomposes the state vector into a spanned component identified from the cross section of yields, and an unspanned latent component that evolves stochastically and is inferred from the dynamic structure of yields. The benchmark specification corresponds to a restricted market-price-of-risk model, henceforth denoted by $M_1$, motivated by prior evidence that unrestricted Gaussian DTSMs often overfit and can perform poorly out of sample. Consistent with the broader evidence in \citet{Bauer18} and related forecasting studies, sparse risk-price restrictions often provide a more robust predictive benchmark than unrestricted alternatives. The latent extension developed here therefore asks whether additional unspanned variation improves forecasting and portfolio performance beyond an already competitive parsimonious model. Our approach is complementary to models based on explicit macroeconomic trend components, capturing related predictive variation directly from yields in a parsimonious latent-state framework.

The proposed specification offers three advantages. First, it avoids reliance on observed macroeconomic covariates and associated real-time data revision issues. Second, the latent component is inferred from yield dynamics through Kalman filtering, allowing persistent predictive variation that is not directly reflected in contemporaneous yields. Third, we develop a sequential Bayesian estimation and forecasting framework based on Sequential Monte Carlo methods. Building on \citet{Chopin02,Chopin04,DelMoral06,Chopin2012} and exploiting the linear Gaussian structure of the model, the procedure combines particle learning for static parameters with Kalman-filter updates for latent states, delivering real-time predictive densities that jointly account for parameter and state uncertainty. The procedure also delivers the full predictive density of bond excess returns in real time, allowing a Bayesian investor to update beliefs and make sequential portfolio decisions.

We evaluate the proposed framework along three dimensions. First, we assess statistical performance using out-of-sample $R^2$ measures following \cite{Campbell08}. Second, we examine economic value through a dynamic portfolio allocation problem with power utility preferences \citep{Gargano19}, computing certainty equivalent returns as in \cite{Johannes14}. Third, inspired by \cite{Duffee11}, we explore the relationship between the latent factor and macroeconomic activity. The empirical results support the proposed framework. Relative to standard benchmark models, the latent extension delivers consistent improvements in out-of-sample return forecasting, particularly at shorter maturities. Although modest in magnitude, the estimated latent factor appears to play a role analogous to the return-forecasting factor in \citet{Cochrane05}, and these gains translate into economically meaningful utility improvements.

As in \citet{Duffee11}, the estimated latent component is countercyclical and linked to real economic activity, suggesting that it captures economically relevant variation not directly observed from yields alone. This behavior is consistent with the idea that shorter-maturity yields are driven by expectations of future policy rates, which are influenced by central bank responses to changing economic conditions \citep{CoroneoGiannoneModugno2016}. The remainder of the paper is organised as follows. Section~\ref{Model} presents the model. Section~\ref{sec:Sequential} describes the sequential learning and forecasting framework. Section~\ref{sec:Data} outlines the data and model specifications. Section~\ref{sec:application} reports the empirical results. Section~\ref{sec:Conclude} concludes.

\section{DTSMs with Unspanned Latent Risks}
\label{Model}

\subsection{Standard Gaussian DTSM setup}
\label{subsec:standardcase}

We begin with the standard no-arbitrage Gaussian affine term structure model and briefly fix notation. More detailed derivations are provided in Section 1 of the Supplementary Material. Let $X_t$ denote a $(N \times 1)$ vector of latent state variables which under the physical measure $\mathbb{P}$ evolve as
\begin{align}
X_{t}&= \mu^{\mathbb{P}} + \Phi^{\mathbb{P}} X_{t-1} + \Sigma \varepsilon_{t}
\end{align}
where $\varepsilon_{t} \sim N(0,\; I_{N})$ and $r_{t}=\delta_{0}+\delta_{1}' X_{t}$ is the short rate. The pricing kernel follows as
\begin{align}
\mathcal{M}_{t+1} 
= \exp\!\left( -r_{t}-\frac{1}{2} \lambda_{t}' \lambda_{t} - \lambda_{t}' \varepsilon_{t+1}\right), \qquad
\lambda_{t} 
= \Sigma^{-1}\left(\lambda_{0} + \lambda_{1} X_{t}\right),
\end{align}
where $\lambda_t$ denotes the vector of market prices of risk. Under standard no-arbitrage arguments, bond prices and yields are affine in $X_t$,
\begin{align}
P_{t}^{n} = \exp(A_{n} + B_{n}' X_{t}), \qquad
y_{t}^{n} = A_{n,X}+B_{n,X}'X_{t},
\end{align}
where $A_{n}$ and $B_{n}$ follow Riccati recursions, and associated risk-neutral dynamics are
\begin{equation}\label{VAR_Q}
X_{t}= \mu^{\mathbb{Q}} + \Phi^{\mathbb{Q}} X_{t-1} + \Sigma \varepsilon_{t}^{\mathbb{Q}},
\end{equation}
where $\mu^{\mathbb{Q}} = \mu^{\mathbb{P}} - \lambda_{0}$ and $\Phi^{\mathbb{Q}} = \Phi^{\mathbb{P}} - \lambda_{1}$. Following \cite{Joslin11}, we rotate the latent state vector to observable factors, using the first $N$ principal components of yields
\begin{align}
y_{t} = A_{\mathcal{P}}+B_{\mathcal{P}} \mathcal{P}_{t}, \label{newyield}
\end{align}
with $\mathcal{P}_{t}=W y_t$, and induced $\mathbb{P}$ and $\mathbb{Q}$ dynamics of the same form. In particular,
\begin{align}
\mu_{\mathcal{P}}^{\mathbb{P}}=\mu_{\mathcal{P}}^{\mathbb{Q}}+\lambda_{0\mathcal{P}}, \qquad
\Phi_{\mathcal{P}}^{\mathbb{P}}=\Phi_{\mathcal{P}}^{\mathbb{Q}}+\lambda_{1\mathcal{P}}.
\end{align}
To account for the fact that an $N$-dimensional state vector cannot perfectly price $J>N$ yields, we allow for measurement error in the remaining yields and write
\begin{equation}
y_{t}=A_{\mathcal{P}}+B_{\mathcal{P}} \mathcal{P}_{t} + e_t,
\label{Qmodel}
\end{equation}
where $e_t$ has variance $\sigma_e^2 I_{J-N}$ in the orthogonal complement of the principal-component space. Finally, for identification we follow \cite{Joslin11} and set
\begin{equation}
r_t = i'X_t, \qquad
\mu^{\mathbb{Q}} = [k_{\infty}^{\mathbb{Q}}, 0, 0]', 
\qquad
\Phi^{\mathbb{Q}} = \mathrm{diag}(g^{\mathbb{Q}}).
\end{equation}

\subsection{Model with unspanned latent components}
\label{subsec:UnspannedLatent}

We extend the baseline Gaussian DTSM by introducing an unspanned latent component that is not directly recoverable from the contemporaneous cross section of yields, yet affects expected return dynamics and predictive content. Our specification is related to \citet{Joslin14}, who augment term structure models with observable macroeconomic factors that are unspanned by yields, and to \citet{Duffee11}, who emphasises hidden return-predictive variation in Gaussian term structure models. Here, we instead model the unspanned component as a latent stochastic process inferred directly from yields. This yields a parsimonious framework that remains entirely yields-based while allowing persistent predictive variation beyond the observed yield curve. It is also complementary to approaches that model persistent trend components through explicit macroeconomic latent states \citep{BauerRudebusch2020}.

A key feature of our specification is a dimension-reduced representation of the spanned space. Specifically, rather than working with all $N$ observable factors, we retain the first $R<N$ principal components and complement them with $(N-R)$ latent components. This yields a decomposition of the state vector into a spanned component $\mathcal{P}_t$ and an unspanned latent component $Z_t$:
\begin{equation}
\label{decomp}
\begin{bmatrix} \mathcal{P}_t \\ Z_t \end{bmatrix}
=
\begin{bmatrix} W_R A_{X} \\ \gamma_0 \end{bmatrix}
+
\begin{bmatrix} W_R B_{X} \\ \gamma_1 \end{bmatrix} X_t,
\end{equation}
where $\mathcal{P}_t$ contains the first $R$ principal components of yields. The matrices $\gamma_0$ and $\gamma_1$ are chosen such that the transformation is invertible (see Appendix C in \cite{Joslin14}). Under this transformation, the physical dynamics of $(\mathcal{P}_t, Z_t)$ take the form
\begin{equation}
\label{Pmodel}
\begin{bmatrix} \mathcal{P}_t \\ Z_t \end{bmatrix}
=
\begin{bmatrix} \mu_{\mathcal{P}}^{\mathbb{P}} \\ \mu_{Z}^{\mathbb{P}} \end{bmatrix}
+
\begin{bmatrix} 
\Phi_{\mathcal{P}}^{\mathbb{P}} & \Phi_{\mathcal{P}Z}^{\mathbb{P}} \\
\Phi_{Z\mathcal{P}}^{\mathbb{P}} & \Phi_{Z}^{\mathbb{P}} 
\end{bmatrix}
\begin{bmatrix} \mathcal{P}_{t-1} \\ Z_{t-1} \end{bmatrix}
+
\Sigma_{\mathcal{P}Z} \varepsilon_t,
\end{equation}
with nonsingular covariance matrix $\Sigma_{\mathcal{P}Z}\Sigma_{\mathcal{P}Z}'$. 

Identification of the latent augmentation is nontrivial because unobserved state components are subject to affine indeterminacy and some unrestricted parameters may be weakly identified in practice. We therefore impose a structured set of economically interpretable restrictions on the loading, transition, and innovation matrices that yields an identified, parsimonious, and tractable specification; formal details are provided in Appendix \ref{appendix:identification}. In particular, we let $\Sigma_{\mathcal{P}Z}\Sigma_{\mathcal{P}Z}'$ be block-diagonal, we set $\mu_Z^{\mathbb{P}} = 0$ and $\Phi_{Z\mathcal{P}}^{\mathbb{P}} = 0$, while letting $\Phi_{\mathcal{P}Z}^{\mathbb{P}}$ be the identity matrix. The latent component $Z_t$ follows a stable autoregressive process with diagonal coefficient $\Phi_Z^{\mathbb{P}}$ and covariance $\Sigma_Z\Sigma_Z'$ matrices. This leads to the following state-space representation:
\begin{align}
\mathcal{P}_t &= \mu_{\mathcal{P}}^{\mathbb{P}} 
+ \Phi_{\mathcal{P}}^{\mathbb{P}} \mathcal{P}_{t-1} 
+ Z_{t-1} 
+ \Sigma_{\mathcal{P}} \varepsilon_t^{\mathcal{P}}, \label{KFobs} \\
Z_t &= \Phi_Z^{\mathbb{P}} Z_{t-1} 
+ \Sigma_Z \varepsilon_t^{Z}, \label{KFstate}
\end{align}
with independent Gaussian innovations.

Relative to the standard model, the latent component enters the drift of the spanned factors as a persistent random effect. This specification captures time-varying deviations from the dynamics implied by the yield curve alone, while preserving tractability.
The latent component $Z_t$ can also be interpreted as a time-varying source of risk premia that is not spanned by the yield curve. In this sense, it plays a role analogous to latent state variables in conditional asset pricing models, capturing systematic variation in expected returns that is not directly observable. Unlike standard factor models, however, $Z_t$ affects returns through the market price of risk while remaining excluded from the cross-sectional pricing relation. Specifically, the pricing recursion is driven by $\mathcal{P}_t$, while the market price of risk is allowed to depend on both $\mathcal{P}_t$ and $Z_t$
\begin{align}
\mathcal{M}_{t+1} 
= \exp\!\left( -r_t - \tfrac{1}{2}\lambda_t'\lambda_t - \lambda_t' \varepsilon_{t+1}^{\mathcal{P}} \right), \qquad
\lambda_t 
= \Sigma_{\mathcal{P}}^{-1}\left( \lambda_{0\mathcal{P}} + \lambda_{1\mathcal{P}} \mathcal{P}_t + Z_t \right),
\end{align}
which ensures that $Z_t$ remains unspanned by construction. As a result, the latent component affects expected returns through the market price of risk, but is not directly recoverable from contemporaneous yields.

The model is completed by specifying the risk-neutral dynamics of $\mathcal{P}_t$ as in \eqref{VAR_Q}. Since $Z_t$ does not enter pricing, its risk-neutral dynamics are not required. This construction provides a parsimonious way to introduce additional predictive state variables without violating the no-arbitrage structure or relying on external covariates.

\subsection{State-space representation and the Kalman filter}
\label{subsec:Kalman}

The model in \eqref{KFobs}–\eqref{KFstate} can be cast into a standard linear Gaussian state-space form (see, e.g., \cite{DurbinKoopman2012}) by defining $\alpha_t = Z_{t-1}$ and specifying the initial condition
\begin{equation}
\alpha_0 \sim N(a_{0|0},\; P_{0|0}), \qquad a_{0|0}=0, \qquad P_{0|0}=\Sigma_Z \Sigma_Z'.
\end{equation}
This representation allows us to marginalise out the latent component $Z_t$ exactly via the Kalman filter and evaluate the conditional densities $f^{\mathbb{P}}(\mathcal{P}_{t}|\mathcal{P}_{t-1},\theta)$. Define
\begin{equation}
s_t = \mathcal{P}_t - \mu_{\mathcal{P}}^{\mathbb{P}} - \Phi_{\mathcal{P}}^{\mathbb{P}} \mathcal{P}_{t-1}.
\label{smalls}
\end{equation}
Then \eqref{KFobs}–\eqref{KFstate} can be rewritten as
\begin{align}
s_t &= \alpha_t + \Sigma_{\mathcal{P}}\varepsilon_{t}^{\mathcal{P}}, \label{KFobsNew}\\
\alpha_{t+1} &= \Phi_{Z}^{\mathbb{P}} \alpha_{t}+ \Sigma_Z \varepsilon_{t}^{Z}. \label{KFstateNew}
\end{align}
Given $(a_{0|0}, P_{0|0})$, the Kalman filter proceeds via the prediction step
\begin{align}
a_{t+1} = \Phi_{Z}^{\mathbb{P}} a_{t|t}, \qquad
P_{t+1} = \Phi_{Z}^{\mathbb{P}} P_{t|t} \Phi_{Z}^{\mathbb{P}'} + \Sigma_Z \Sigma_Z',
\end{align}
and the update step is
\begin{align}
v_t &= s_t - a_t, \qquad
F_t = P_t + \Sigma_{\mathcal{P}}\Sigma_{\mathcal{P}}', \qquad
K_t = P_t F_t^{-1}, \\
a_{t|t} &= a_t + K_t v_t, \qquad
P_{t|t} = P_t - K_t P_t. \label{statevar}
\end{align}
Here $v_t$ denotes the prediction error, $F_t$ its covariance, and $K_t$ the Kalman gain. Consequently, the latent state $\alpha_t$ (equivalently $Z_{t-1}$) is distributed as $N(a_{t|t}, P_{t|t})$. The Kalman filter yields the Gaussian log-likelihood contribution (see \cite{Schweppe1965})
\begin{equation}
\log f^{\mathbb{P}}(\mathcal{P}_{t}|\mathcal{P}_{t-1}, \theta) 
= -\frac{R}{2}\log(2\pi)
-\frac{1}{2}\left(\log|F_t| + v_t' F_t^{-1} v_t\right),
\label{loglikP}
\end{equation}
which provides a tractable likelihood after integrating out the latent component. Finally, with $\alpha_T \sim N(a_{T|T}, P_{T|T})$, the predictive distribution of $\mathcal{P}_{T+1}$ is given by
\begin{equation}
\mathcal{P}_{T+1}|\mathcal{P}_{T},\alpha_T 
\sim N\left(
\mu_{\mathcal{P}}^{\mathbb{P}} 
+ \Phi_{\mathcal{P}}^{\mathbb{P}}\mathcal{P}_{T}
+ \Phi_{Z}^{\mathbb{P}} \alpha_{T},
\;
\Sigma_Z\Sigma_Z' + \Sigma_\mathcal{P}\Sigma_\mathcal{P}'
\right),
\label{predictiveP}
\end{equation}

\subsection{Likelihood and risk price restrictions}
\label{subsec:likelihood}

Statistical inference is based on the observations $Y=\{y_t,\mathcal{P}_{t}: t=0,1,\dots,T\}$. The likelihood naturally factorises into a cross-sectional component under the risk-neutral measure $\mathbb{Q}$ and a time-series component under the physical measure $\mathbb{P}$. For $R$ observable factors, the joint likelihood (conditional on $\mathcal{P}_{0}$) is given by
\begin{align}
f(Y|\theta,\widehat{\Sigma}_Z)
&=
\prod_{t=0}^T 
f^{\mathbb{Q}}\!\left(y_{t}|\mathcal{P}_{t}, k_{\infty}^{\mathbb{Q}}, g^{\mathbb{Q}}, \Sigma_{\mathcal{P}}, \sigma_{e}^{2}\right)
\nonumber \\
&\quad \times
\prod_{t=1}^T 
f^{\mathbb{P}}\!\left(\mathcal{P}_{t}|\mathcal{P}_{t-1}, k_{\infty}^{\mathbb{Q}}, g^{\mathbb{Q}}, \lambda_{0\mathcal{P}}, \lambda_{1\mathcal{P}}, \Sigma_{\mathcal{P}}, \Phi_{Z}^{\mathbb{P}}, \Sigma_Z \right),
\label{likelihood}
\end{align}
where the $\mathbb{Q}$-likelihood parts capture the cross-sectional fit of yields (see \eqref{Qmodel}), while the $\mathbb{P}$-likelihood parts are obtained via the Kalman filter (see \eqref{loglikP}) and capture the time-series dynamics after integrating out the latent component. The parameter vector is
\[
\theta=(\sigma_{e}^{2},k_{\infty}^{\mathbb{Q}}, g^{\mathbb{Q}},\lambda_{0\mathcal{P}},\lambda_{1\mathcal{P}}, \Sigma_{\mathcal{P}}, \Phi_{Z}^{\mathbb{P}}, \Sigma_Z).
\]
In practice, the latent innovation variance proved weakly identified. We therefore calibrate $\Sigma_Z$ to a data-driven estimate $\widehat{\Sigma}_Z$ obtained from in-sample information under a restricted yields-only specification; details are provided in Section 4 of the Supplementary Material. This anchors the scale of the latent process to observed yield dynamics, promoting stable inference while preserving flexibility in the latent component.

\medskip
\noindent\textbf{Risk Price Restrictions and Benchmark Models.} For notational convenience, let $\lambda^{\mathcal{P}}=[\lambda_{0\mathcal{P}},\lambda_{1\mathcal{P}}]$ and $\lambda=\lambda_{1\mathcal{P}}$. We consider two benchmark specifications. Model $M_0$ leaves all elements of $\lambda^{\mathcal{P}}$ unrestricted. Model $M_1$ imposes a sparse structure in which only $\lambda_{1,2}$ is freely estimated and all remaining elements are set to zero. The restricted specification is motivated by evidence that fully flexible Gaussian DTSMs are often weakly identified, unstable, and prone to overfitting out of sample; see, among others, \citet{Bauer18}. In related forecasting work, we selected restrictions using a sequential stochastic search variable-selection procedure with emphasis on out-of-sample predictive and economic performance \citep{DTKK2024}. This yielded the parsimonious $M_1$ specification adopted here. Related findings in \citet{Duffee02,Duffee11} also emphasize time variation in the price of level risk, often linked to the slope of the term structure.

Under these restrictions, and after fixing $\Sigma_Z$ to $\widehat{\Sigma}_Z$, the likelihood in \eqref{likelihood} simplifies to
\begin{align}
f(Y|\theta,\widehat{\Sigma}_Z)
&=
\prod_{t=0}^T 
f^{\mathbb{Q}}\!\left(y_{t}|\mathcal{P}_{t}, k_{\infty}^{\mathbb{Q}}, g^{\mathbb{Q}}, \Sigma_{\mathcal{P}}, \sigma_{e}^{2}\right)
\nonumber \\
&\quad \times
\prod_{t=1}^T 
f^{\mathbb{P}}\!\left(\mathcal{P}_{t}|\mathcal{P}_{t-1}, k_{\infty}^{\mathbb{Q}}, g^{\mathbb{Q}}, \Sigma_{\mathcal{P}}, \Phi_{Z}^{\mathbb{P}}, \lambda_{1,2}, \widehat{\Sigma}_Z \right),
\label{likelihoodlambda12}
\end{align}
with parameter vector
\[
\theta=(\sigma_{e}^{2},k_{\infty}^{\mathbb{Q}}, g^{\mathbb{Q}},\Sigma_{\mathcal{P}}, \Phi_{Z}^{\mathbb{P}},\lambda_{1,2}).
\]
This likelihood formulation is particularly convenient for sequential learning: the latent component is integrated analytically through the Kalman filter, while the remaining static parameters are updated using particle methods.

\section{Sequential Estimation, Filtering, and Forecasting}
\label{sec:Sequential}

This section develops the sequential computational framework used for estimation, filtering and forecasting in the proposed model. We first present an Iterated Batch Importance Sampling (IBIS) scheme adapted to a DTSM with unspanned latent components, where Kalman-filter updates are embedded within each particle. We then describe how the resulting particle approximation is used to construct predictive distributions of excess bond returns and to assess their statistical and economic value.

The proposed inference framework builds on the IBIS approach of \cite{DTKK2024}, originally developed for yields-only term structure models with directly observed state variables. In the present setting, the introduction of unspanned latent factors requires joint inference on both parameters and latent states. To address this, we integrate Kalman filtering within the IBIS scheme, allowing for sequential updating of parameter distributions while propagating latent states through the state-space representation.

\subsection{Sequential Framework with Latent Processes}

Let $Y_{0:t}=(Y_0,Y_1,\dots,Y_t)$ denote the data available up to time $t$, so that $Y_{0:T}=Y$. The corresponding likelihood based on data up to time $t$ is $f(Y_{0:t}|\theta,\widehat{\Sigma}_Z)$, as defined in \eqref{likelihood}. Combined with a prior $\pi(\theta)$, which is specified in Appendix \ref{appendix:priors}, this yields the posterior
\begin{equation}\label{posterior}
 \pi(\theta|Y_{0:t},\widehat{\Sigma}_Z)= \frac{1}{m(Y_{0:t}|\widehat{\Sigma}_Z)}f(Y_{0:t}|\theta,\widehat{\Sigma}_Z)\pi(\theta),
\end{equation}
where $m(Y_{0:t}|\widehat{\Sigma}_Z)$ denotes the model evidence. The corresponding posterior predictive distribution is
\begin{equation}\label{predictive}
f(Y_{t+h}\mid Y_{0:t},\widehat{\Sigma}_Z)
=
\iint
f(Y_{t+h}\mid Y_t,\alpha_t,\theta,\widehat{\Sigma}_Z)\,
\pi(\alpha_t\mid Y_{0:t},\theta,\widehat{\Sigma}_Z)\,
\pi(\theta\mid Y_{0:t},\widehat{\Sigma}_Z)\,d\alpha_t\,d\theta,
\end{equation}
where $h$ denotes the prediction horizon. Conditional on $\theta$, the filtering distribution of $\alpha_t$ is Gaussian and delivered by the Kalman filter, allowing the integration over $\alpha_t$ to be performed analytically. The key methodological contribution is the integration of IBIS with exact Kalman-filter updates for latent states, enabling joint sequential inference on parameters and unobserved components without resorting to particle filtering for the state process. This yields an efficient sequential scheme tailored to the state-space structure of the model.

The predictive distribution in \eqref{predictive} integrates out parameter uncertainty through the posterior in \eqref{posterior}. In principle, one could approximate \eqref{predictive} separately at each forecast origin using Monte Carlo draws from $\pi(\theta|Y_{0:t},\widehat{\Sigma}_Z)$. In practice, this becomes computationally demanding when forecasts must be evaluated repeatedly over a long out-of-sample period. To address this issue, we use sequential Monte Carlo to sample from the sequence of posteriors $\pi(\theta|Y_{0:t},\widehat{\Sigma}_Z)$ for $t=0,1,\dots,T$. We build on the Iterated Batch Importance Sampling (IBIS) scheme of \cite{Chopin02}; see also \cite{DelMoral06}. The resulting algorithm is adapted to our setting by combining particle learning for the static parameters with Kalman updates for the latent state. A generic description is given in Algorithm \ref{tab:ibis}.

\begin{algorithm}[!ht]
\begin{flushleft}
{\itshape \small
\vspace{0.2cm}
Initialise $N_{\theta}$ particles by drawing independently $[\theta_{i},\alpha_{0}^{(i)}]\sim [\pi(\theta),N(a_{0|0},P_{0|0})]$ with importance weights $\omega_{i}=1$, $i=1,\dots,N_{\theta}$. For $t=1,\dots,T$ and for all $i$:\vspace{0.2cm}
\begin{itemize}
\item[(a)] Compute incremental weights
$$
u_t([\theta_{i},\alpha_{t-1}^{(i)}]) =f\big(Y_{t}|Y_{t-1},[\theta_{i},\alpha_{t-1}^{(i)}],\widehat{\Sigma}_Z\big),
$$
where $\alpha_{t-1}^{(i)}\sim N\left(a_{t-1|t-1}^{(i)},\;P_{t-1|t-1}^{(i)}\right)$. Update $\alpha_{t-1}^{(i)}$ to $\alpha_t^{(i)}\sim N\left(a_{t|t}^{(i)},\;P_{t|t}^{(i)}\right)$ using the Kalman filter.

\item[(b)] Update the importance weights from $\omega_i$ to $\omega_i\,u_t([\theta_{i},\alpha_{t-1}^{(i)}])$.

\item[(c)] If a degeneracy criterion is triggered:
\begin{itemize}
\item[(i)] Resample: sample with replacement $N_{\theta}$ times from the particles $\theta_i$ according to weights $\omega_i$, and reset the weights to one.
\item[(ii)] Jitter: replace $\theta_i$ with $\tilde{\theta}_i$ by running short MCMC chains, jointly updating the Kalman-filtered latent state. Set $[\theta_i,\alpha_t^{(i)}]=[\tilde{\theta}_i,\widetilde{\alpha}_t^{(i)}]$.
\end{itemize}
\end{itemize}}
\end{flushleft}
\caption{IBIS algorithm for Gaussian affine term structure models with unspanned latent factors}
\label{tab:ibis}
\end{algorithm}
While the core building blocks of the algorithm—sequential importance sampling, resampling, and particle rejuvenation—follow standard IBIS implementations, their use in the present setting requires adaptation. In particular, likelihood evaluations are obtained via the Kalman filter, and particle rejuvenation steps are constructed to account for the resulting filtered likelihood. Degeneracy is monitored using the Effective Sample Size
\begin{equation} \label{ESS}
ESS(\omega)=\frac{(\sum_{i=1}^{N_{\theta}} \omega_{i})^2}{\sum_{i=1}^{N_{\theta}} \omega_{i}^2},
\end{equation}
and resampling is triggered whenever $ESS(\omega)<\alpha N_{\theta}$ for some $\alpha \in (0,1)$.

The IBIS output provides a weighted particle approximation to the posterior at each time $t$, and can therefore be used to compute expectations of the form $E[g(\theta)|Y_{0:t},\widehat{\Sigma}_Z]$ through
\[
\frac{\sum_i \omega_i g(\theta_i)}{\sum_i \omega_i}.
\]
By \cite{Chopin04}, this estimator is consistent and asymptotically normal as $N_{\theta}\to\infty$ for suitably integrable $g(\cdot)$. The same weighted sample can also be propagated through the predictive mechanism to approximate posterior predictive distributions. A further by-product of IBIS is sequential estimation of the model evidence. In particular, the quantity
\begin{equation}
m_t = \frac{1}{\sum_{i=1}^{N_{\theta}}\omega_{i}}\sum_{i=1}^{N_{\theta}}\omega_{i}u_{t}([\theta_{i},\alpha_{t-1}^{(i)}])
\end{equation}
provides a consistent and asymptotically normal estimator of $f(Y_t|Y_{0:t-1},\widehat{\Sigma}_Z)$.

To tailor IBIS to the model proposed in this paper, three modifications are required. First, the latent process is updated sequentially within each particle via Kalman filtering, avoiding the need for a separate particle filter. Second, to stabilise learning, we employ a hybrid tempering scheme combining data tempering and adaptive tempering, following \cite{DTKK2024}. In the present setting, tempering is applied to likelihood contributions derived from the Kalman filter, allowing for smoother incorporation of new observations and improved robustness in periods of heightened volatility. Particle rejuvenation is implemented using independence Metropolis--Hastings updates based on local approximations of the posterior, following \cite{DTKK2024}. In contrast to the yields-only case, the proposals are evaluated using likelihood contributions obtained via the Kalman filter. This affects both acceptance probabilities and mixing behaviour; details are given in Sections 2 and 3 of the Supplementary Material. Third, since the MCMC rejuvenation step relies on independence samplers extending the approach in \citet{Bauer18}, we use posterior moments estimated from the particle cloud to construct proposal distributions adaptively.

These modifications are essential for obtaining stable and reliable inference in the presence of unspanned latent components. They ensure computational feasibility, enable joint learning of parameters and latent states, and provide a practical route to sequential prediction in the proposed framework.

\subsection{Assessing Predictive Performance and Economic Value}
\label{sec:predEconomicValue}

We next use the sequential particle approximation to assess whether unspanned latent information improves the prediction of bond excess returns relative to corresponding yields-only models, and whether any statistical gains translate into economic value for investors.

\subsubsection{Bond Excess Returns}
\label{sec:R2OS}

Define the observed $h$-period holding return from buying an $n$-year bond at time $t$ and selling it at time $t+h$ as
\begin{equation}\label{holding}
r_{t,t+h}^{n} = p_{t+h}^{n-h} - p_{t}^{n},
\end{equation}
where $p_{t+h}^{n-h}$ is the log price of the $(n-h)$-period bond at time $t+h$ and $p_t^n$ is the log price of the $n$-period bond at time $t$. Since $p_t^n=-n y_t^n$, the observed continuously compounded excess return is
\begin{equation}\label{excess}
rx_{t,t+h}^{n} = -(n-h) y_{t+h}^{n-h} + ny_{t}^{n} - hy_{t}^{h}.
\end{equation}
Using model-implied yields from \eqref{newyield} provides the corresponding predicted excess return
\begin{equation}\label{fitted} 
\widetilde{rx}_{t,t+h}^{n}
= A_{n-h,\mathcal{P}} - A_{n,\mathcal{P}} + A_{h,\mathcal{P}} + B'_{n-h,\mathcal{P}} \widetilde{\mathcal{P}}_{t+h} - (B_{n,\mathcal{P}}-B_{h,\mathcal{P}})'\mathcal{P}_{t},
\end{equation}
where $\mathcal{P}_t$ is observed and $\widetilde{\mathcal{P}}_{t+h}$ is generated from the predictive distribution of the model.

The proposed framework combines two complementary elements. First, the introduction of unspanned latent components allows the model to capture persistent variation in the conditional mean of the spanned factors that is not directly observable from the yield curve. Second, the IBIS--Kalman scheme enables sequential learning about both parameters and latent states, allowing the model to adapt over time as new data arrive. Together, these features yield predictive distributions that incorporate both hidden state dynamics and parameter uncertainty, providing a flexible basis for forecasting and decision-making. Denote the particle approximation to the predictive distribution, provided by the IBIS--Kalman scheme, by $(\widetilde{\mathcal{P}}_{t+h},\widetilde{rx}_{t,t+h}^{n})$ conditional on information available at time $t$. For each particle $\theta_i$, the $\mathbb{P}$-dynamics of the model yield a draw of $\widetilde{\mathcal{P}}_{t+h}$, which can then be mapped into a draw of $\widetilde{rx}_{t,t+h}^{n}$ through \eqref{fitted}. Detailed steps for the case $h=1$ are summarised in Algorithm \ref{tab:predictive}.

\begin{algorithm}[!ht]
\begin{flushleft}
{\itshape \small
\vspace{0.2cm}
At time $t$, for some maturity $n$ and horizon $h=1$, use the weighted particle sample $(\omega_i,\theta_i)$, $i=1,\dots,N_\theta$:

\begin{itemize}
\item[(a)] For each $\theta_i$, compute $A_{m,\mathcal{P}}$ and $B_{m,\mathcal{P}}$ for $m\in\{1,n-1,n\}$.

\item[(b)] Draw
$$
\widetilde{\alpha}_{t+1}^{(i)}|\alpha_t^{(i)} \sim N\left(\Phi_{Z}^{\mathbb{P}} \alpha_t^{(i)},\; \widehat{\Sigma}_Z\widehat{\Sigma}_Z'\right),
$$
and then
$$
\widetilde{\mathcal{P}}_{t+1}^{(i)}|\mathcal{P}_{t},\widetilde{\alpha}_{t+1}^{(i)} 
\sim N\left(\mu_{\mathcal{P}}^{\mathbb{P}} + \Phi_{\mathcal{P}}^{\mathbb{P}}\mathcal{P}_{t} + \widetilde{\alpha}_{t+1}^{(i)},\; \Sigma_\mathcal{P}\Sigma_\mathcal{P}'\right),
$$
where $\alpha_t^{(i)}\sim N(a_{t|t}^{(i)},P_{t|t}^{(i)})$.

\item[(c)] Compute the particle prediction
$$
\widetilde{rx}_{t,t+1}^{n(i)} = A_{n-1,\mathcal{P}} - A_{n,\mathcal{P}} + A_{1,\mathcal{P}} + B'_{n-1,\mathcal{P}} \widetilde{\mathcal{P}}_{t+1}^{(i)} - (B_{n,\mathcal{P}}-B_{1,\mathcal{P}})'\mathcal{P}_{t}.
$$
\end{itemize}

The point prediction is then obtained as the weighted average
$$
\widetilde{rx}_{t,t+1}^{n}
=\frac{1}{\sum_{i=1}^{N_\theta}\omega_{i}}\sum_{i=1}^{N_{\theta}}\omega_{i}\widetilde{rx}_{t,t+1}^{n(i)}.
$$

For $h>1$, the same procedure applies recursively, replacing $\mathcal{P}_t$ and $\alpha_t^{(i)}$ by their simulated values at intermediate horizons.}
\end{flushleft}
\caption{Predictive distribution of excess returns for Gaussian affine term structure models with unspanned latent factors}
\label{tab:predictive}
\end{algorithm}
To assess predictive performance, we first adopt the Expectations Hypothesis benchmark, which uses historical averages as forecasts of bond excess returns:
\begin{equation}\label{eh}
\overline{rx}_{t+h}^{n} = \frac{1}{t-h} \sum_{j=1}^{t-h} rx_{j,j+h}^{n}.
\end{equation}
We then evaluate forecasts using the out-of-sample $R^2$ measure of \cite{Campbell08},
\begin{equation}\label{R2os}
R_{os}^{2} = 1 - \frac{\sum_{s=t_{0}}^{t} ( rx_{s,s+h}^{n} -\widetilde{rx}_{s,s+h}^{n})^{2}}{\sum_{s=t_{0}}^{t} (rx_{s,s+h}^{n} -\overline{rx}_{s+h}^{n})^{2}},
\end{equation}
where $\widetilde{rx}_{s,s+h}^{n}$ denotes the mean of the predictive distribution. Positive values indicate that the model-implied forecasts outperform the benchmark. To gauge the magnitude of the gains relative to the EH or model $M_1$, we also report one-sided Diebold--Mariano statistics with the \cite{ClarkWest2007} adjustment, following \cite{WanFulopLi2022}. As in the related literature, these are interpreted informally as indices rather than formal hypothesis tests.

\subsubsection{Economic Performance of Excess Return Forecasts}
\label{sec:EconomicValue}

From the perspective of a bond investor, it is also important to assess whether predictive gains translate into economically meaningful portfolio benefits out-of-sample. To do so, we follow \cite{Gargano19} and consider a Bayesian investor with power utility
\begin{equation}\label{utility}
U(W_{t+h}) = \frac{W_{t+h}^{1-\gamma}}{1-\gamma},
\end{equation}
where $W_{t+h}$ is the $h$-period portfolio value and $\gamma$ is the coefficient of relative risk aversion. Let $w_t^n$ denote the portfolio weight invested in the risky $n$-period bond, with the remaining share $(1-w_t^n)$ invested in the risk-free $h$-period bond. The resulting portfolio value is
\begin{equation}\label{portfolio}
W_{t+h} = (1-w_{t}^{n}) \exp (r_{t}^{f}) + w_{t}^{n} \exp(r_{t}^{f} + rx_{t,t+h}^{n}),
\end{equation}
where $r_t^f$ denotes the risk-free rate.

Given the predictive density of excess returns, the investor chooses $w_t^n$ to maximise expected utility. Using the particle approximation from the IBIS algorithm, the optimal weight is obtained numerically as
\begin{equation}
\widetilde{w}_{t}^{n} = \arg \max_{w_t^n}\,
\frac{1}{\sum_{i=1}^{N_\theta} \omega_{i}} \sum_{i=1}^{N_{\theta}} \omega_{i}
\left\{
\frac{[(1-w_{t}^{n})\exp(r_{t}^{f}) + w_{t}^{n}\exp(r_{t}^{f} + \widetilde{rx}_{t,t+h}^{n,i})]^{1-\gamma}}{1-\gamma}
\right\}.
\end{equation}
To measure economic value, we compute certainty equivalent returns (CER), following \cite{Johannes14} and \cite{Gargano19}. Letting $\widetilde{U}_t$ denote realised utility under the predictive model and $\overline{U}_t$ the corresponding utility under the EH benchmark, CER is defined by
\begin{equation}\label{CER}
CER = \left(\frac{\sum_{s=t_0}^t \widetilde{U}_s}{\sum_{s=t_0}^t \overline{U}_s}\right)^{\frac{1}{1-\gamma}}-1.
\end{equation}
We report CER values relative to both the EH benchmark and model $M_1$. The portfolio allocations implied by the model remain within economically reasonable bounds across all specifications. As in the case of $R_{os}^2$, statistical significance is summarised using one-sided Diebold--Mariano statistics with Newey--West adjusted standard errors, interpreted informally as indices rather than formal tests.

\section{Data and Models}
\label{sec:Data}

This section describes the yield and macroeconomic data used in the empirical analysis, and the model specifications considered in the forecasting and economic value exercises.

\subsection{Yields and Macros}
\label{subsec:yieldsdata}

The yield data set contains monthly observations of US Treasury zero-coupon bond yields\footnote{The yields are unsmoothed Fama--Bliss yields constructed by \cite{Le13}. We thank Ahn Le for generously providing the data set.} with maturities of $1$, $2$, $3$, $4$, $5$, $7$ and $10$ years, spanning the period from January 1985 to December 2018. The sample is split into a training period ending in December 2007 and a testing period running from January 2008 to December 2018. Excluding the post-2007 financial crisis period from the training sample allows the models to learn about that episode sequentially. This is particularly relevant in our setting, since the testing period includes unconventional monetary policy and extended periods near the zero lower bound, where the performance of Gaussian DTSMs may be more challenging; see, for example, \cite{Kim12,Bauer16}. Data processing involves extracting principal components from the yield curve. In the empirical analysis we use the first three principal components, with loadings estimated from the training sample only, in order to avoid look-ahead bias. Using loadings estimated from the full sample leads to negligible differences, with correlations above $0.99$ between the resulting series.

To help interpret the latent components ex post, we also consider a small set of monthly US macroeconomic variables. These include core inflation ($CPI$), following \cite{Cieslak15}, the three-month moving average of the Chicago Fed National Activity Index ($GRO$), following \cite{Joslin14}, and the real activity ($F1$), cubic real activity ($F1^3$), and stock market ($F8$) factors of \cite{Ludvigson09}. We complement these with the unemployment rate ($UNR$) and manufacturing capacity utilisation ($MNF$).\footnote{The latter two series are obtained from the St.\ Louis FRED database.} All macroeconomic series are observed at monthly frequency over a period aligned with the yield data.

\subsection{Models}
\label{subsec:modelsrationale}

In addition to the yields-only benchmark models $M_0$ and $M_1$, we consider a family of models obtained by augmenting $M_1$ with unspanned latent components. These models differ according to the positions at which the latent components enter the state dynamics. We label them $LF_{ijk}$, $i,j,k\in\{0,1\}$, where, for example, $LF_{011}$ denotes a specification in which latent components enter the second and third equations of \eqref{KFobs}, under $R=3$.

Under the restrictions defining model $M_1$, the stochastic risk premium factor is
\begin{equation}\label{RPZ111}
RP^Z_t 
=
\begin{bmatrix}
\lambda_{1,2}\mathcal{P}_{2,t} + Z_{1,t} \\
Z_{2,t} \\
Z_{3,t}
\end{bmatrix},
\end{equation}
where the first element is a restricted version of the time-varying risk premium factor in \cite{Duffee11}, since only $\lambda_{1,2}$ is left unrestricted. In this specification, investors are compensated for level risk through time variation linked to the slope factor, while the latent component $Z_{1,t}$ adds an unspanned stochastic element to that compensation. If instead $i=0$ and $j=k=1$, the first element coincides with the time-varying risk premium factor of the yields-only model $M_1$, and we obtain
\begin{equation}\label{RPZ011}
RP^Z_t =
\begin{bmatrix}
\lambda_{1,2}\mathcal{P}_{2,t} \\
Z_{2,t} \\
Z_{3,t}
\end{bmatrix}.
\end{equation}

These alternative specifications allow us to assess whether the latent stochastic component contributes information relevant for predicting excess returns and generating economic value beyond that already contained in the yields-only benchmark. They also allow us to examine whether the latent factors remain hidden from the yield curve, and whether any such hidden component is related to macroeconomic activity. To that end, and following the empirical design of \cite{Duffee11}, we define the hidden part of the stochastic risk premium factor as the component orthogonal to $\mathcal{P}_t$:
\begin{align}
\widetilde{RP}^Z_t 
&= RP^Z_t - E\left[RP^Z_t|\mathcal{P}_t\right] = Z_t - E[Z_t|\mathcal{P}_t]. \label{RPZhiddenshort}
\end{align}
Here $E[Z_t|\mathcal{P}_t]$ is the projection of the latent component on the principal components extracted from observed yields and therefore corresponds to the spanned part. 
The hidden component is assessed ex post by removing from the latent factor the part linearly explained by the observed principal components. In the baseline implementation, we use the posterior mean path of $Z_t$,  $\widehat{Z}_t=E(Z_t\mid Y)$, and define
\begin{equation}\label{RPZreg}
\widehat{\widetilde{RP}}{}^Z_t
=
\widehat{Z}_t - \hat a - \hat b'\mathcal{P}_t,
\end{equation}
where $(\hat a,\hat b)$ are ordinary least squares estimates. This decomposition is intended as a posterior-based diagnostic of hidden variation rather than a separate Bayesian estimand.

Finally, we provide some implementation details. In the empirical application presented in Section~\ref{sec:application}, we use $N_{\theta}=2,000$ particles, $5$ MCMC steps at each jittering stage, and set the minimum ESS threshold to $\alpha=0.7$. Results were qualitatively unchanged under larger particle sets. The choice of $5$ MCMC steps reflects the satisfactory mixing behaviour of the underlying sampler, as assessed through particle correlations before and after the rejuvenation step. Results are robust to moderate variations in the number of particles and MCMC steps; the chosen configuration reflects a trade-off between computational cost and numerical stability.

\section{Empirical Results}
\label{sec:application}

This section presents the statistical and economic performance of the models introduced earlier, based on the IBIS--Kalman framework developed in Section \ref{sec:Sequential}. We assess whether unspanned latent factors improve the predictability of bond excess returns and whether such gains translate into economic value.

\subsection{Observing the Unobserved}

\begin{figure}[t]
\centering
\includegraphics[trim = 22mm 25mm 0mm 0mm, height=4.4in,left]{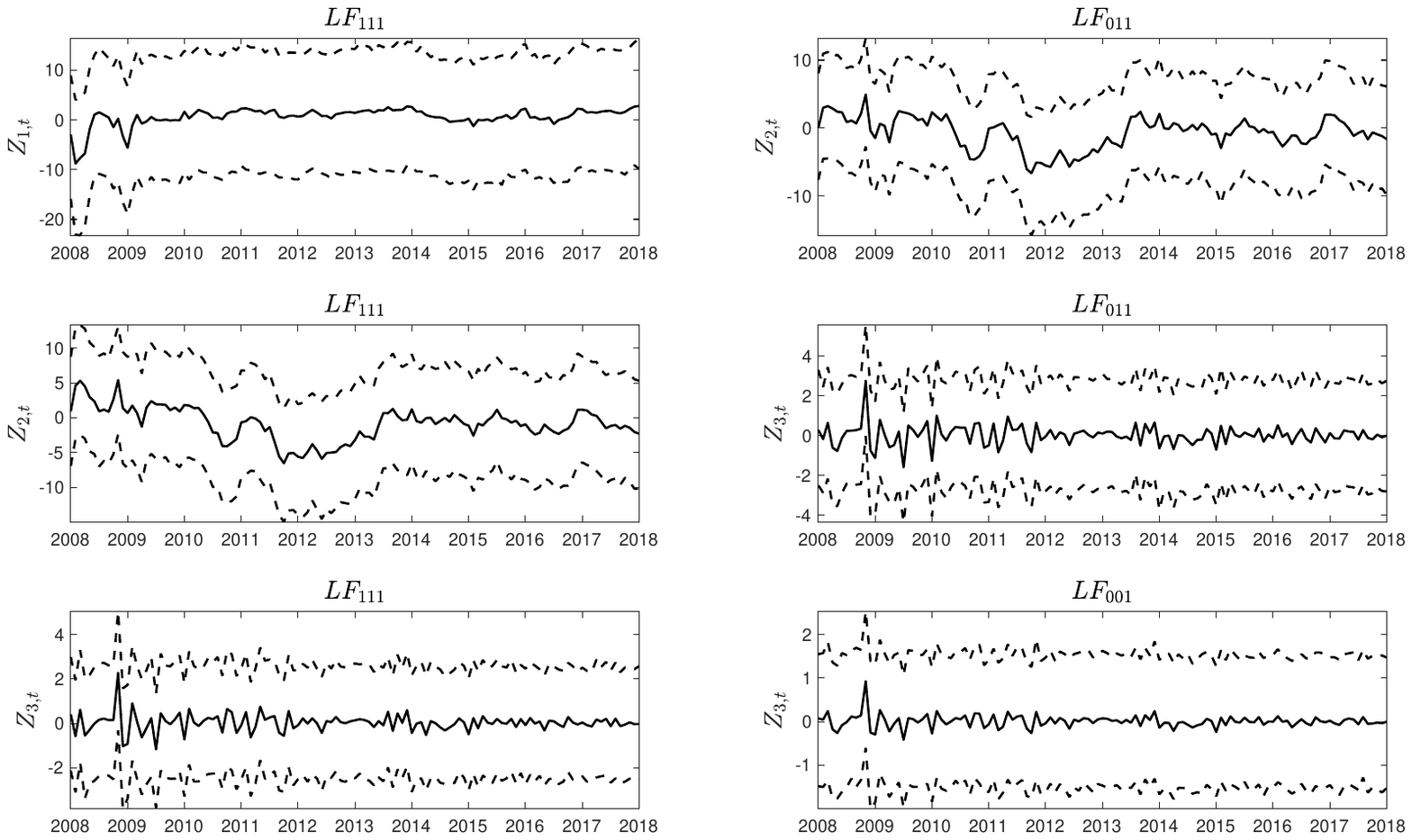}
\caption[Unspanned latent factors filtered from the yield curve]{Unspanned latent factors filtered from the yield curve in basis points per annum. The figure focuses on the period from January 2008 to December 2017, over which predictive performance is also evaluated, although the latent factors are estimated using the full sample from January 1985 onward. The left column presents the factors from model $LF_{111}$. The right column shows the factors from models $LF_{011}$ (first and second row) and $LF_{001}$ (third row). Solid lines represent posterior means and dashed lines the 95\% credible intervals.}
\label{fig:Zs}
\end{figure}

The estimated latent factors exhibit clear and economically interpretable dynamics around major events, such as the 2008--2009 financial crisis (Figure \ref{fig:Zs}). In particular, the second latent factor $Z_{2,t}$ displays a pronounced and persistent decline during the crisis period, consistent with shifts in slope-related risks, while $Z_{1,t}$ and $Z_{3,t}$ capture shorter-lived and more oscillatory deviations associated with level and curvature dynamics.

Overall, the latent factors capture variation aligned with key movements in the yield curve, while also introducing additional dynamics not directly spanned by observed yields.

\subsection{Bond Return Predictability and Economic Performance}

\subsubsection{Bond Return Predictability}
\label{sec:bondreturnpred}

Table \ref{table:R2OS} reports out-of-sample $R^2$ values for one-month ahead forecasts over the period 2008--2018, using 1985--2007 data as the estimation sample. Relative to the Expectations Hypothesis (EH) benchmark, most models achieve positive $R^2_{os}$ across maturities, indicating meaningful return predictability. The exception is the fully flexible model $M_0$, which performs poorly out-of-sample. Models incorporating unspanned latent factors generally deliver improved predictive performance, particularly at shorter maturities. For example, model $LF_{010}$ achieves $R^2_{os}$ of 5.86\% for the 2-year bond and 3.42\% for the 10-year bond, with similar results for $LF_{011}$. These gains highlight the importance of slope-related risks captured by the second latent factor.

\begin{table}[t] \scriptsize
\caption[Out-of-sample statistical performance of bond excess return forecasts against the EH and yields-only model $M_1$]{Out-of-sample statistical performance of bond excess return forecasts against the EH and model $M_1$, measured via $R_{os}^2$ ($\%$) at $h=1$ month, with the training sample running from January 1985 to December 2007 and the evaluation period from January 2008 to December 2018.}
\label{table:R2OS}
\centering
\begin{tabular}{lllllll}
\toprule
\multicolumn{1}{m{3.0cm}}{$\bf h=1m$ $\backslash$ $\bf n$} &\multicolumn{1}{m{0.5cm}}{\bf 2Y} &\multicolumn{1}{m{0.5cm}}{\bf 3Y} & \multicolumn{1}{m{0.5cm}}{\bf 4Y} & \multicolumn{1}{m{0.5cm}}{\bf 5Y} & \multicolumn{1}{m{0.5cm}}{\bf 7Y} & \multicolumn{1}{m{1cm}}{\bf 10Y}\\
\midrule
\multicolumn{7}{c}{\bf Panel A: forecasts against the EH benchmark} \\
\midrule
\bf $M_{0}$ & -3.84 & -5.54 & -4.74 & -3.60 & -1.54 & -1.45** \\
\bf $M_{1}$ & 1.30 & 2.90** & 2.52* & 1.98 & 2.49* & 3.86** \\
\bf $LF_{001}$ & 2.91** & 4.35** & 3.64** & 3.22* & 3.48* & 4.70*** \\
\bf $LF_{010}$ & 5.86*** & 6.12*** & 4.61*** & 3.71** & 3.00** & 3.42*** \\
\bf $LF_{011}$ & 5.65*** & 5.30*** & 3.87*** & 2.79** & 2.10* & 2.57** \\
\bf $LF_{100}$ & 2.52* & 4.00** & 2.58* & 2.16 & 2.63 & 4.03* \\
\bf $LF_{110}$ & 5.56*** & 5.30*** & 3.13** & 2.40* & 2.11* & 2.56** \\
\bf $LF_{111}$ & 5.05*** & 4.53*** & 2.84** & 2.08* & 1.52* & 2.16** \\
\midrule
\multicolumn{7}{c}{\bf Panel B: forecasts against yields-only model $M_{1}$} \\
\midrule
\bf $LF_{001}$ & 1.62* & 1.49*** & 1.15** & 1.26** & 1.01** & 0.87* \\
\bf $LF_{010}$ & 4.62*** & 3.31*** & 2.14*** & 1.77*** & 0.53 & -0.46 \\
\bf $LF_{011}$ & 4.40*** & 2.47*** & 1.38** & 0.83 & -0.40 & -1.34 \\
\bf $LF_{100}$ & 1.24 & 1.13 & 0.06 & 0.18 & 0.15 & 0.18 \\
\bf $LF_{110}$ & 4.31** & 2.47* & 0.62 & 0.42 & -0.39 & -1.35 \\
\bf $LF_{111}$ & 3.80** & 1.68 & 0.32 & 0.11 & -1.00 & -1.77 \\
\bottomrule
\end{tabular}
\caption*{\tiny This table reports out-of-sample $R^2$ across alternative models at the one-month prediction horizon. Panel A presents forecasts relative to the EH benchmark, while Panel B presents forecasts relative to the yields-only model $M_1$. Positive values indicate improved predictive accuracy. Statistical significance is measured using a one-sided Diebold--Mariano statistic with Clark--West adjustment, based on Newey--West standard errors. * denotes significance at 10\%, ** at 5\%, and *** at 1\%.}
\end{table}

When compared to the best-performing yields-only model $M_1$ (Panel B), models with latent factors continue to outperform, indicating that unspanned components contain predictive information beyond that embedded in the yield curve. Improvements are again most pronounced at shorter maturities, suggesting that the additional information is associated with the short end of the maturity spectrum.

The strong performance of models incorporating the second latent factor suggests that time variation in slope-related risks plays a central role in driving excess bond returns. This is consistent with the view that deviations from the Expectations Hypothesis arise from time-varying risk premia linked to the term structure slope. The latent component $Z_{2,t}$ appears to capture additional variation in this risk premium that is not fully reflected in observed yields, providing a complementary source of predictive information beyond the cross section of interest rates.

\subsubsection{Economic Performance}

We next evaluate whether statistical predictability translates into economic gains for investors. Table \ref{table:CER} reports certainty equivalent returns (CER) across three investment scenarios with varying portfolio constraints. The first two prevent investors from taking extreme positions, while the third relaxes restrictions and allows for maximum leveraging and short-selling. Positive values indicate that models perform better than the EH benchmark. Across all scenarios, models with unspanned latent factors consistently deliver positive and statistically significant CER gains relative to the benchmark. The gains are particularly pronounced for models $LF_{001}$ and $LF_{010}$ and tend to increase as portfolio constraints are relaxed.

\begin{table}[t]\scriptsize
\caption[Out-of-sample economic performance of bond excess return forecasts against the EH]{Out-of-sample economic performance of bond excess return forecasts against the EH, measured via certainty equivalent returns (\%) at $h=1$ month, with the training sample running from January 1985 to December 2007 and the evaluation period from January 2008 to December 2018.}
\label{table:CER}
\centering
\begin{tabular}{lllllll}
\toprule
\multicolumn{1}{m{1.5cm}}{$\bf h=1m \backslash \bf n$} &\multicolumn{1}{m{0.5cm}}{\bf 2Y} &\multicolumn{1}{m{0.5cm}}{\bf 3Y} & \multicolumn{1}{m{0.5cm}}{\bf 4Y} & \multicolumn{1}{m{0.5cm}}{\bf 5Y} & \multicolumn{1}{m{0.5cm}}{\bf 7Y} & \multicolumn{1}{m{1cm}}{\bf 10Y}\\
\midrule
\multicolumn{7}{c}{\bf Panel A: $w\in[-1,2]$} \\
\midrule
\bf $M_{0}$ & -1.09 & -1.61 & -2.43 & -2.63 & -2.09 & -4.46 \\
\bf $M_{1}$ & 0.00 & 0.00 & -0.04 & 0.10 & 1.23** & 0.78 \\
\bf $LF_{001}$ & 0.00 & 0.00 & 0.09* & 0.13 & 1.36** & 1.85* \\
\bf $LF_{010}$ & 0.00 & 0.00 & 0.12** & 0.37* & 1.49** & 1.38 \\
\bf $LF_{011}$ & 0.00 & 0.00 & 0.10* & 0.34* & 1.24** & 0.76 \\
\bf $LF_{100}$ & 0.00 & 0.07 & 0.19 & 0.23 & 1.34* & 0.75 \\
\bf $LF_{110}$ & 0.14 & 0.23* & 0.28* & 0.23 & 1.13** & 0.93 \\
\bf $LF_{111}$ & 0.17 & 0.28* & 0.41* & 0.61* & 1.63** & 0.87 \\
\midrule
\multicolumn{7}{c}{\bf Panel B: $w\in[-1,5]$} \\
\midrule
\bf $M_{0}$ & -2.04 & -3.52 & -3.38 & -3.20 & -1.82 & -3.85 \\
\bf $M_{1}$ & -0.09 & 0.02 & 0.80 & 0.46 & 1.32 & 1.51 \\
\bf $LF_{001}$ & 0.08* & 0.16 & 1.40* & 1.83 & 2.96* & 2.85** \\
\bf $LF_{010}$ & 0.19* & 0.42 & 1.54* & 1.83 & 2.48* & 1.84* \\
\bf $LF_{011}$ & 0.21* & 0.35 & 1.68** & 1.50 & 2.08 & 1.25 \\
\bf $LF_{100}$ & 0.23 & 0.40 & 0.81 & -0.07 & -0.02 & 0.11 \\
\bf $LF_{110}$ & 0.25 & 0.22 & 0.77 & 0.74 & 1.59 & 0.24 \\
\bf $LF_{111}$ & 0.47* & 0.57 & 1.10 & 0.15 & 1.10 & 0.17 \\
\midrule
\multicolumn{7}{c}{\bf Panel C: no portfolio bounds} \\
\midrule
\bf $M_{0}$ & -6.37 & -7.05 & -6.13 & -4.82 & -2.16 & -4.50 \\
\bf $M_{1}$ & 1.42 & 1.16 & 0.83 & 0.48 & 1.32 & 1.51 \\
\bf $LF_{001}$ & 2.42** & 2.31** & 2.12* & 1.96 & 2.96* & 2.85** \\
\bf $LF_{010}$ & 2.40** & 2.31** & 2.15** & 1.83 & 2.48 & 1.84* \\
\bf $LF_{011}$ & 2.26** & 2.01* & 1.90* & 1.47 & 2.08 & 1.25 \\
\bf $LF_{100}$ & 0.22 & 0.35 & -0.20 & -0.42 & -0.02 & 0.11 \\
\bf $LF_{110}$ & 0.61 & 0.99 & 0.80 & 0.75 & 1.59 & 0.24 \\
\bf $LF_{111}$ & 0.02 & 0.28 & 0.25 & 0.13 & 1.10 & 0.17 \\
\bottomrule
\end{tabular}
\caption*{\tiny This table reports annualised certainty equivalent returns (CERs) across alternative models at the one-month prediction horizon. The coefficient of relative risk aversion is $\gamma=5$. Panels A--C correspond to increasingly loose portfolio constraints. Positive values indicate that the models perform better than the EH benchmark. Statistical significance is measured using a one-sided Diebold--Mariano statistic with Newey--West standard errors. * denotes significance at 10\%, ** at 5\%, and *** at 1\%.}
\end{table}

In contrast, the yields-only benchmark $M_1$ generates limited or insignificant economic value, while the fully flexible model $M_0$ performs poorly out-of-sample. These findings indicate that unspanned latent factors enhance not only statistical predictability but also economic performance. When evaluated relative to $M_1$ (Table \ref{table:CERM1}), the gains remain positive but more moderate, suggesting that latent factors provide incremental economic value beyond the best-performing yields-only specification. The improvements are strongest for models that include latent factors associated with slope and curvature dynamics, while models placing latent factors on the level component perform less well.

\begin{table}[t]\scriptsize
\caption[Out-of-sample economic performance of bond excess return forecasts against model $M_1$]{Out-of-sample economic performance of bond excess return forecasts against model $M_1$, measured via certainty equivalent returns (\%) at $h=1$ month -- period: January 1985 -- end of 2018.}
\label{table:CERM1}
\centering
\begin{tabular}{lllllll}
\toprule
\multicolumn{1}{m{1.5cm}}{$\bf h=1m \backslash \bf n$} &\multicolumn{1}{m{0.5cm}}{\bf 2Y} &\multicolumn{1}{m{0.5cm}}{\bf 3Y} & \multicolumn{1}{m{0.5cm}}{\bf 4Y} & \multicolumn{1}{m{0.5cm}}{\bf 5Y} & \multicolumn{1}{m{0.5cm}}{\bf 7Y} & \multicolumn{1}{m{1cm}}{\bf 10Y}\\
\midrule
\multicolumn{7}{c}{\bf Panel A: $w\in[-1,2]$} \\
\midrule
\bf $LF_{001}$ & 0.00 & 0.00 & 0.13** & 0.03 & 0.13 & 1.07** \\
\bf $LF_{010}$ & 0.00 & 0.00 & 0.16** & 0.27** & 0.26 & 0.60 \\
\bf $LF_{011}$ & 0.00 & 0.00 & 0.14** & 0.23** & 0.00 & -0.03 \\
\bf $LF_{100}$ & 0.00 & 0.07 & 0.23 & 0.12 & 0.10 & -0.04 \\
\bf $LF_{110}$ & 0.14 & 0.23* & 0.32* & 0.13 & -0.10 & 0.14 \\
\bf $LF_{111}$ & 0.17 & 0.28* & 0.45** & 0.51** & 0.40 & 0.08 \\
\midrule
\multicolumn{7}{c}{\bf Panel B: $w\in[-1,5]$} \\
\midrule
\bf $LF_{001}$ & 0.17* & 0.14 & 0.59* & 1.37** & 1.64*** & 1.34** \\
\bf $LF_{010}$ & 0.27** & 0.40* & 0.74** & 1.37*** & 1.17** & 0.33 \\
\bf $LF_{011}$ & 0.30** & 0.33 & 0.88** & 1.04* & 0.76 & -0.26 \\
\bf $LF_{100}$ & 0.32* & 0.38 & 0.01 & -0.53 & -1.33 & -1.39 \\
\bf $LF_{110}$ & 0.33 & 0.20 & -0.03 & 0.28 & 0.28 & -1.27 \\
\bf $LF_{111}$ & 0.56* & 0.55 & 0.29 & -0.31 & -0.22 & -1.33 \\
\midrule
\multicolumn{7}{c}{\bf Panel C: no portfolio bounds} \\
\midrule
\bf $LF_{001}$ & 1.00*** & 1.15*** & 1.28** & 1.48*** & 1.64*** & 1.34** \\
\bf $LF_{010}$ & 0.97* & 1.15** & 1.31*** & 1.35** & 1.17** & 0.33 \\
\bf $LF_{011}$ & 0.83* & 0.85* & 1.07** & 0.98* & 0.76 & -0.26 \\
\bf $LF_{100}$ & -1.20 & -0.81 & -1.03 & -0.90 & -1.33 & -1.39 \\
\bf $LF_{110}$ & -0.81 & -0.17 & -0.03 & 0.26 & 0.28 & -1.27 \\
\bf $LF_{111}$ & -1.40 & -0.88 & -0.58 & -0.36 & -0.22 & -1.33 \\
\bottomrule
\end{tabular}
\caption*{\tiny This table reports annualised certainty equivalent returns (CERs) across alternative models at the one-month prediction horizon, relative to the yields-only benchmark model $M_1$. Panels A--C correspond to increasingly loose portfolio constraints. Positive values indicate improved economic performance relative to $M_1$. Statistical significance is measured using a one-sided Diebold--Mariano statistic with Newey--West standard errors. * denotes significance at 10\%, ** at 5\%, and *** at 1\%.}
\end{table}

Overall, the results provide strong evidence that unspanned latent risks contain economically meaningful information that is not captured by the cross section of yields alone. This is a notable finding, especially because investors use only yield curve information and no additional macroeconomic inputs, unlike studies that draw predictive information from macroeconomic sources; see, for example, \citet{Gargano19}, \citet{Bianchi20}, and \citet{Wan21}.

\subsection{Linking Unspanned Latent Risks with the Macroeconomy}

We next investigate whether the hidden component of the latent risk-premium factors is related to macroeconomic conditions. In particular, we focus on the component of the risk premium that is orthogonal to the yield curve, as defined in \eqref{RPZhiddenshort}.

Table \ref{table:R2adjRPZEZPZMACROS1} reports adjusted $R^2$ values from regressions of the hidden component on macroeconomic variables. Results indicate that a significant fraction of the variation in the slope-related hidden factor ($\widetilde{RP}^Z_{2,t}$) is explained by real activity indicators. For example, measures such as $GRO$, $MNF$, and $UNR$ explain between 11\% and 27\% of its variation across specifications.

\begin{table}[t]\scriptsize
\caption[Explanatory power of macroeconomic variables when fitting latent factors and their components]{Explanatory power of macroeconomic variables when fitting latent factors and their components, measured via $\bar{R}^2$ -- period: January 1985 -- end of 2017.}
\label{table:R2adjRPZEZPZMACROS1}
\centering
\begin{tabular}{c|llllllllll}
\toprule
\multicolumn{11}{c}{$\bar{R}^2:Z_{j,t}/E[Z_{j,t}|\mathcal{P}_t]/\widetilde{RP}^Z_{j,t}=a_j+b_j'M_t+e_{j,t},\;j\in\{1,2,3\}$}\\
\midrule 
\multicolumn{1}{c|}{$LF_{001}$} & \multicolumn{1}{l}{$CPI$} & \multicolumn{1}{l}{$GRO$} & \multicolumn{1}{l}{${F1}$} & \multicolumn{1}{l}{${F1^3}$} & \multicolumn{1}{l}{${F8}$} & \multicolumn{1}{l}{$UNR$} &\multicolumn{1}{l}{$MNF$} & \multicolumn{1}{c}{$M^I$} & \multicolumn{1}{c}{$M^{II}$} & \multicolumn{1}{c}{$M^{III}$} \\
\midrule
$Z_{3,t}$  &       0.00 &       0.00 &       0.00 &       0.00 &       0.00 &       0.00 &       0.00 &       0.00 &       0.00 &       0.00 \\
\\
$E[Z_{3,t}|\mathcal{P}_t]$  &       0.00 &       0.17 &       0.17 &       0.14 &       0.00 &       0.04 &       0.00 &       0.17 &       0.18 &       0.06 \\
$\widetilde{RP}^Z_{3,t}$  &       0.00 &       0.01 &       0.01 &       0.01 &       0.00 &       0.00 &       0.00 &       0.00 &       0.01 &       0.00 \\
\midrule 
\multicolumn{1}{c|}{$LF_{010}$} & \multicolumn{1}{l}{$CPI$} & \multicolumn{1}{l}{$GRO$} & \multicolumn{1}{l}{${F1}$} & \multicolumn{1}{l}{${F1^3}$} & \multicolumn{1}{l}{${F8}$} & \multicolumn{1}{l}{$UNR$} &\multicolumn{1}{l}{$MNF$} & \multicolumn{1}{c}{$M^I$} & \multicolumn{1}{c}{$M^{II}$} & \multicolumn{1}{c}{$M^{III}$} \\
\midrule
$Z_{2,t}$  &       0.05 &       0.03 &       0.03 &       0.00 &       0.02 &       0.20 &       0.08 &       0.08 &       0.05 &       0.21 \\
\\
$E[Z_{2,t}|\mathcal{P}_t]$  &       0.09 &       0.01 &       0.00 &       0.00 &       0.01 &       0.06 &       0.00 &       0.11 &       0.01 &       0.17 \\
$\widetilde{RP}^Z_{2,t}$  &       0.00 &       0.11 &       0.09 &       0.02 &       0.01 &       0.15 &       0.19 &       0.11 &       0.10 &       0.20 \\
\midrule 
\multicolumn{1}{c|}{$LF_{011}$} & \multicolumn{1}{l}{$CPI$} & \multicolumn{1}{l}{$GRO$} & \multicolumn{1}{l}{${F1}$} & \multicolumn{1}{l}{${F1^3}$} & \multicolumn{1}{l}{${F8}$} & \multicolumn{1}{l}{$UNR$} &\multicolumn{1}{l}{$MNF$} & \multicolumn{1}{c}{$M^I$} & \multicolumn{1}{c}{$M^{II}$} & \multicolumn{1}{c}{$M^{III}$} \\
\midrule
$Z_{2,t}$  &       0.06 &       0.01 &       0.02 &       0.00 &       0.02 &       0.19 &       0.07 &       0.07 &       0.04 &       0.20 \\
\\
$E[Z_{2,t}|\mathcal{P}_t]$  &       0.09 &       0.02 &       0.01 &       0.01 &       0.01 &       0.04 &       0.00 &       0.12 &       0.02 &       0.13 \\
$\widetilde{RP}^Z_{2,t}$  &       0.00 &       0.13 &       0.11 &       0.03 &       0.01 &       0.18 &       0.21 &       0.13 &       0.12 &       0.23 \\
\midrule
$Z_{3,t}$  &       0.00 &       0.00 &       0.00 &       0.00 &       0.00 &       0.00 &       0.00 &       0.00 &      -0.01 &       0.00 \\
\\
$E[Z_{3,t}|\mathcal{P}_t]$  &       0.00 &       0.17 &       0.16 &       0.14 &       0.00 &       0.03 &       0.00 &       0.17 &       0.18 &       0.03 \\
$\widetilde{RP}^Z_{3,t}$  &       0.00 &       0.01 &       0.01 &       0.01 &       0.00 &       0.00 &       0.01 &       0.01 &       0.01 &       0.01 \\
\midrule 
\multicolumn{1}{c|}{$LF_{111}$} & \multicolumn{1}{l}{$CPI$} & \multicolumn{1}{l}{$GRO$} & \multicolumn{1}{l}{${F1}$} & \multicolumn{1}{l}{${F1^3}$} & \multicolumn{1}{l}{${F8}$} & \multicolumn{1}{l}{$UNR$} &\multicolumn{1}{l}{$MNF$} & \multicolumn{1}{c}{$M^I$} & \multicolumn{1}{c}{$M^{II}$} & \multicolumn{1}{c}{$M^{III}$} \\
\midrule
$Z_{1,t}$  &       0.05 &       0.12 &       0.17 &       0.01 &       0.00 &       0.16 &       0.15 &       0.18 &       0.21 &       0.18 \\
\\
$E[Z_{1,t}|\mathcal{P}_t]$  &       0.10 &       0.00 &       0.00 &       0.00 &       0.02 &       0.09 &       0.00 &       0.10 &       0.01 &       0.24 \\
$\widetilde{RP}^Z_{1,t}$  &       0.02 &       0.14 &       0.17 &       0.01 &       0.00 &       0.11 &       0.17 &       0.16 &       0.22 &       0.18 \\
\midrule
$Z_{2,t}$  &       0.07 &       0.04 &       0.05 &       0.01 &       0.01 &       0.26 &       0.09 &       0.11 &       0.07 &       0.26 \\
\\
$E[Z_{2,t}|\mathcal{P}_t]$  &       0.09 &       0.01 &       0.00 &       0.00 &       0.01 &       0.06 &       0.00 &       0.10 &       0.01 &       0.17 \\
$\widetilde{RP}^Z_{2,t}$  &       0.00 &       0.18 &       0.17 &       0.04 &       0.00 &       0.23 &       0.27 &       0.18 &       0.18 &       0.29 \\
\midrule
$Z_{3,t}$  &       0.00 &       0.00 &       0.00 &       0.00 &       0.00 &       0.00 &       0.01 &       0.00 &       0.00 &       0.01 \\
\\
$E[Z_{3,t}|\mathcal{P}_t]$  &       0.00 &       0.18 &       0.17 &       0.14 &       0.00 &       0.04 &       0.00 &       0.18 &       0.19 &       0.05 \\
$\widetilde{RP}^Z_{3,t}$  &       0.00 &       0.02 &       0.03 &       0.02 &       0.00 &       0.01 &       0.01 &       0.02 &       0.03 &       0.01 \\
\bottomrule
\end{tabular}
\caption*{\tiny This table reports in-sample $\bar{R}^2$ across alternative regression specifications. The explained variables are individual latent factors $Z_{j,t}$ and their components $E[Z_{j,t}|\mathcal{P}_t]$ and $\widetilde{RP}^Z_{j,t}$, $j\in\{1,2,3\}$, from models $LF_{001}$, $LF_{010}$, $LF_{011}$ and $LF_{111}$. The explanatory variables are individual macroeconomic variables or groups thereof, namely $M^I=[CPI,GRO]'$, $M^{II}=[F1,F1^3,F8]'$, and $M^{III}=[UNR,MNF]'$.}
\end{table}

Moreover, the hidden component of the slope-related risk premium is found to be countercyclical, as indicated by negative (for $GRO$, $MNF$ and $F1$), or positive (for $UNR$), and statistically significant coefficients on real activity variables in Table \ref{table:bRPZEZPZMACROS1}. This suggests that shocks to economic activity are associated with systematic variation in bond risk premia that is not fully captured by the yield curve.

\begin{table}[t]\scriptsize
\caption[Signs and significance of coefficients from explanatory power regressions of latent factors and their components on macroeconomic variables]{Signs and significance of coefficients from explanatory power regressions of latent factors and their components on macroeconomic variables -- period: January 1985 -- end of 2017.}
\label{table:bRPZEZPZMACROS1}
\centering
\begin{tabular}{c|lllllll}
\toprule
\multicolumn{8}{c}{$sign(b_j):Z_{j,t}/E[Z_{j,t}|\mathcal{P}_t]/\widetilde{RP}^Z_{j,t}=a_j+b_j M_t+e_{j,t},\;j\in\{1,2,3\}$}\\
\midrule 
\multicolumn{1}{c|}{$LF_{001}$} & \multicolumn{1}{l}{$CPI$} & \multicolumn{1}{l}{$GRO$} & \multicolumn{1}{l}{${F1}$} & \multicolumn{1}{l}{${F1^3}$} & \multicolumn{1}{l}{${F8}$} & \multicolumn{1}{l}{$UNR$} &\multicolumn{1}{l}{$MNF$}\\
\midrule
$Z_{3,t}$  &        $-$ &        $-$ &        $-$ &        $-$ &        $+$ &        $+$ &        $+$ \\
$E[Z_{3,t}|\mathcal{P}_t]$  &        $-$ &  $-^{***}$ &  $-^{***}$ &  $-^{***}$ &        $+$ &    $+^{*}$ &        $-$ \\
$\widetilde{RP}^Z_{3,t}$  &        $-$ &        $+$ &        $+$ &    $+^{*}$ &        $+$ &        $-$ &        $+$ \\
\midrule 
\multicolumn{1}{c|}{$LF_{010}$} & \multicolumn{1}{l}{$CPI$} & \multicolumn{1}{l}{$GRO$} & \multicolumn{1}{l}{${F1}$} & \multicolumn{1}{l}{${F1^3}$} & \multicolumn{1}{l}{${F8}$} & \multicolumn{1}{l}{$UNR$} &\multicolumn{1}{l}{$MNF$}\\
\midrule
$Z_{2,t}$  &   $+^{**}$ &   $-^{**}$ &   $-^{**}$ &    $-^{*}$ &  $+^{***}$ &  $+^{***}$ &  $-^{***}$ \\
$E[Z_{2,t}|\mathcal{P}_t]$  &  $+^{***}$ &   $+^{**}$ &        $+$ &  $+^{***}$ &   $+^{**}$ &   $+^{**}$ &        $+$ \\
$\widetilde{RP}^Z_{2,t}$  &        $+$ &  $-^{***}$ &  $-^{***}$ &  $-^{***}$ &    $+^{*}$ &  $+^{***}$ &  $-^{***}$ \\
\midrule 
\multicolumn{1}{c|}{$LF_{011}$} & \multicolumn{1}{l}{$CPI$} & \multicolumn{1}{l}{$GRO$} & \multicolumn{1}{l}{${F1}$} & \multicolumn{1}{l}{${F1^3}$} & \multicolumn{1}{l}{${F8}$} & \multicolumn{1}{l}{$UNR$} &\multicolumn{1}{l}{$MNF$}\\
\midrule
$Z_{2,t}$  &  $+^{***}$ &    $-^{*}$ &   $-^{**}$ &        $-$ &   $+^{**}$ &  $+^{***}$ &  $-^{***}$ \\
\\
$E[Z_{2,t}|\mathcal{P}_t]$  &  $+^{***}$ &  $+^{***}$ &   $+^{**}$ &  $+^{***}$ &   $+^{**}$ &   $+^{**}$ &        $+$ \\
$\widetilde{RP}^Z_{2,t}$  &        $+$ &  $-^{***}$ &  $-^{***}$ &  $-^{***}$ &    $+^{*}$ &  $+^{***}$ &  $-^{***}$ \\
\midrule
$Z_{3,t}$  &        $-$ &        $-$ &        $-$ &        $-$ &        $+$ &        $-$ &    $+^{*}$ \\
\\
$E[Z_{3,t}|\mathcal{P}_t]$  &        $-$ &  $-^{***}$ &  $-^{***}$ &  $-^{***}$ &        $-$ &        $+$ &        $-$ \\
$\widetilde{RP}^Z_{3,t}$  &        $-$ &   $+^{**}$ &    $+^{*}$ &   $+^{**}$ &        $+$ &    $-^{*}$ &   $+^{**}$ \\
\midrule 
\multicolumn{1}{c|}{$LF_{111}$} & \multicolumn{1}{l}{$CPI$} & \multicolumn{1}{l}{$GRO$} & \multicolumn{1}{l}{${F1}$} & \multicolumn{1}{l}{${F1^3}$} & \multicolumn{1}{l}{${F8}$} & \multicolumn{1}{l}{$UNR$} &\multicolumn{1}{l}{$MNF$}\\
\midrule
$Z_{1,t}$  &  $-^{***}$ &  $+^{***}$ &  $+^{***}$ &  $+^{***}$ &        $+$ &  $-^{***}$ &  $+^{***}$ \\
\\
$E[Z_{1,t}|\mathcal{P}_t]$  &  $-^{***}$ &        $-$ &        $-$ &        $-$ &   $-^{**}$ &  $-^{***}$ &        $-$ \\
$\widetilde{RP}^Z_{1,t}$  &   $-^{**}$ &  $+^{***}$ &  $+^{***}$ &  $+^{***}$ &   $+^{**}$ &  $-^{***}$ &  $+^{***}$ \\
\midrule
$Z_{2,t}$  &  $+^{***}$ &  $-^{***}$ &  $-^{***}$ &   $-^{**}$ &   $+^{**}$ &  $+^{***}$ &  $-^{***}$ \\
\\
$E[Z_{2,t}|\mathcal{P}_t]$  &  $+^{***}$ &   $+^{**}$ &        $+$ &  $+^{***}$ &   $+^{**}$ &   $+^{**}$ &        $+$ \\
$\widetilde{RP}^Z_{2,t}$  &        $+$ &  $-^{***}$ &  $-^{***}$ &  $-^{***}$ &        $+$ &  $+^{***}$ &  $-^{***}$ \\
\midrule
$Z_{3,t}$  &   $-^{**}$ &        $+$ &        $+$ &        $+$ &        $+$ &        $-$ &   $+^{**}$ \\
\\
$E[Z_{3,t}|\mathcal{P}_t]$  &        $-$ &  $-^{***}$ &  $-^{***}$ &  $-^{***}$ &        $+$ &        $+$ &        $-$ \\
$\widetilde{RP}^Z_{3,t}$  &        $-$ &  $+^{***}$ &  $+^{***}$ &   $+^{**}$ &        $+$ &   $-^{**}$ &   $+^{**}$ \\
\bottomrule
\end{tabular}
\caption*{\tiny This table reports signs and statistical significance of coefficients $b_j$, $j\in\{1,2,3\}$, across alternative regression specifications. The explained variables are individual latent factors $Z_{j,t}$ and their components $E[Z_{j,t}|\mathcal{P}_t]$ and $\widetilde{RP}^Z_{j,t}$, $j\in\{1,2,3\}$, from models $LF_{001}$, $LF_{010}$, $LF_{011}$ and $LF_{111}$. The explanatory variables are macroeconomic variables. Statistical significance is measured using Newey--West adjusted $t$-statistics. * denotes significance at 10\%, ** at 5\%, and *** at 1\%.}
\end{table}

Taken together, the results highlight the value of combining structured no-arbitrage models with flexible latent components and sequential estimation methods. This provides a unified framework in which methodological advances translate into economically meaningful improvements in prediction and decision-making.


\section{Conclusions}
\label{sec:Conclude}

This paper studies whether information hidden from the yield curve can improve predictions of bond excess returns and deliver economically meaningful gains to bond investors. We develop an arbitrage-free DTSM with unspanned latent factors, combining a dimension-reduced representation of the yield curve with a stochastic component that captures variation not directly reflected in yields. Estimation and forecasting are conducted using a sequential Monte Carlo scheme coupled with Kalman filtering, allowing for real-time learning about both parameters and latent states.

Empirically, we find evidence of out-of-sample return predictability relative to the EH benchmark. Models with unspanned latent factors consistently outperform the best-performing yields-only specification, indicating that these factors contain predictive information beyond the cross section of yields. The gains are particularly pronounced at shorter maturities, suggesting that unspanned components capture variation relevant to the short end of the term structure. Importantly, these statistical improvements translate into economically meaningful gains. Models with unspanned latent risks generate positive and significant certainty equivalent returns across a range of portfolio constraints, and continue to outperform the yields-only benchmark. This indicates that the additional information extracted by the latent factors is not only statistically relevant but also valuable for investment decisions.

Finally, we document a link between unspanned latent risks and macroeconomic conditions. The hidden component of the slope-related factor is closely related to real activity indicators and exhibits countercyclical behaviour. This suggests that the latent factors capture economically meaningful variation in risk premia that is not fully reflected in the yield curve. Overall, our results highlight the value of augmenting standard term structure models with unspanned latent components. By combining an arbitrage-free structure with flexible latent dynamics and sequential learning, the proposed framework provides a tractable approach to extracting and exploiting information that is otherwise hidden from the cross section of yields.

\section*{Data Availability Statement}
The data and code that support the findings of this study are available from the corresponding author upon reasonable request. The data were obtained from a third-party provider and processed using the methodology described in the manuscript.

\FloatBarrier

\bibliography{referencefile3.bib}
\newpage
\phantomsection\label{appendix}
\bigskip

\begin{center}

{\LARGE\bf Appendix
}

\end{center}

\begin{appendices}

\section{Econometric Identification of the Model}
\label{appendix:identification}

If $Z_t$ were observed, then under the normalisation adopted for the spanned factors $\mathcal{P}_t$ following \citet{Joslin11}, the model would be identified as in \citet{Joslin14}. In the present setting, however, $Z_t$ is latent, so additional restrictions are required. The physical dynamics of $(\mathcal{P}_t,Z_t)$ are initially given by
\begin{equation}
\label{Pmodelapp}
\begin{bmatrix} \mathcal{P}_t \\ Z_t \end{bmatrix}
=
\begin{bmatrix} \mu_{\mathcal{P}}^{\mathbb{P}} \\ \mu_{Z}^{\mathbb{P}} \end{bmatrix}
+
\begin{bmatrix}
\Phi_{\mathcal{P}}^{\mathbb{P}} & \Phi_{\mathcal{P}Z}^{\mathbb{P}} \\
\Phi_{Z\mathcal{P}}^{\mathbb{P}} & \Phi_{Z}^{\mathbb{P}}
\end{bmatrix}
\begin{bmatrix} \mathcal{P}_{t-1} \\ Z_{t-1} \end{bmatrix}
+
\Sigma_{\mathcal{P}Z}\varepsilon_t,
\end{equation}
with covariance matrix
\[
\Sigma_{\mathcal{P}Z}\Sigma_{\mathcal{P}Z}'=
\begin{bmatrix}
\Sigma_{\mathcal{P}}\Sigma_{\mathcal{P}}' & C_{Z\mathcal{P}} \\
C_{Z\mathcal{P}}' & \Sigma_Z\Sigma_Z'
\end{bmatrix}.
\]
Because $Z_t$ is unobserved, the model is invariant to affine transformations of the form
\begin{equation}
\tilde Z_t=\Gamma_0+\Gamma_1 Z_t,
\label{Ztransform}
\end{equation}
for conformable constant matrices $\Gamma_0$ and nonsingular $\Gamma_1$. Absent further restrictions, this generates observationally equivalent parameterisations.

\medskip
\noindent\textbf{Proposition A1.} Under the latent-state representation \eqref{Pmodelapp}, imposing
(i) $\mu_Z^{\mathbb P}=0$,
(ii) $\Phi_{\mathcal P Z}^{\mathbb P}=I$,
(iii) diagonal $\Phi_Z^{\mathbb P}$,
(iv) diagonal $\Sigma_Z$, and
(v) $C_{Z\mathcal P}=0$,
pins down the affine normalization in \eqref{Ztransform} and yields an identified representation of the model.\\
\ \\
\textit{Justification:} Condition (i) fixes the location of the latent state. Conditions (ii)--(iv) remove scale and rotation indeterminacy while yielding a parsimonious and economically interpretable specification in which each latent component loads directly on a corresponding observed factor. Condition (v) rules out contemporaneous innovation correlation between spanned and unspanned shocks.

These restrictions are also motivated by interpretation. Throughout the empirical analysis, the first three principal components of yields retain their standard level, slope, and curvature interpretation. Setting $\Phi_{\mathcal P Z}^{\mathbb P}=I$ preserves a transparent mapping between latent and observed factors, while diagonal latent dynamics imply that the unspanned components evolve autonomously a priori. 

\section{Specification of Priors}
\label{appendix:priors}

This section briefly describes prior distributions not detailed in the main text. For parameters in $g^{\mathbb{Q}}$, $k_{\infty}^{\mathbb{Q}}$, $\sigma_e^2$ and $\lambda^{\mathcal{P}}$, or effectively $\lambda_{1,2}$, priors are constructed in the same manner as in the Online Appendix B.1 to \cite{DTKK2024}. The only exceptions are parameters in $\Phi_Z$ and $\Sigma_{\mathcal{P}}$.

For both remaining parameter blocks, we first transform constrained components to have support on the real line. We also rescale these of their elements which
typically take very small values. Specifically, we work with the \textit{logit} transformation to
constrain diagonal elements of $\Phi_Z$ to $(-1,1)$ and consider a Cholesky factorisation of $\Sigma_{\mathcal{P}}$ where the diagonal elements are transformed to the real line and off-diagonal elements are scaled by $10^4$. Next, independent normal distributions with zero means
are assigned to each of their components. Independent zero-mean normal priors are then assigned to all transformed components. Large prior variances are used for elements of $\Sigma_{\mathcal P}$, while the transformed diagonal elements of $\Phi_Z$ are assigned variance $2$.
\end{appendices}

\end{document}